\documentclass{jpp}
\usepackage{xcolor}
\usepackage[utf8]{inputenc}
\usepackage[T1]{fontenc}
\usepackage{comment}
\usepackage{epstopdf, epsfig}
\usepackage{hyperref}
\usepackage{amsmath,amsfonts,amssymb}
\usepackage{graphicx}
\usepackage[normalem]{ulem}
\usepackage{subcaption}

\newcommand{\un}[1]{\,\mathrm{#1}}
\title{Role of the edge electric field in the resonant mode-particle interactions and the formation of transport barriers in toroidal plasmas}

\author{Giorgos Anastassiou\aff{1}
  \corresp{\email{ganastas@central.ntua.gr}}, Panagiotis Zestanakis\aff{1}, Yiannis Antonenas\aff{1}, Eleonora Viezzer\aff{2}, and Yannis Kominis\aff{1}}

\affiliation{\aff{1}School of Applied Mathematical \& Physical Sciences, \\
National Technical University of Athens, Athens, Greece
\aff{2}Department of Atomic, Molecular and Nuclear Physics, University of Seville, Seville, Spain}

\begin{document}

%Exclude graphics for faster compilation
%\excludecomment{figure} 
%\let\endfigure\relax

\maketitle

\begin{abstract}
The impact of an edge radial electric field on the particle orbits and the orbital spectrum in an axisymmetric toroidal magnetic equilibrium is investigated using a guiding center  canonical formalism. Poloidal and bounce/transit-averaged toroidal precession frequencies are calculated, highlighting the role of the radial electric field. The radial electric field is shown to drastically modify the resonance conditions between particles with certain kinetic characteristics and specific perturbative non-axisymmetric modes and to enable the formation of transport barriers. The locations of the resonances and the transport barriers, that determine the particle, energy and momentum transport are shown to be accurately pinpointed in the phase space, by employing the calculated orbital frequencies.
\end{abstract}

\section{Introduction}
Resonant mode-particle interactions have a crucial role in determining particle, energy and momentum transport and confinement in fusion plasmas. Resonant interactions take place at specific locations of the phase space and affect particles with certain kinetic characteristics defined by their energy, momentum and pitch angle \citep{Kaufman1972, Escande1985, Horton1990}, rendering the phase space of the particle motion  strongly inhomogeneous in terms of the effect of a perturbative mode. In general, resonant mode-particle interactions strongly perturb the integrability of the particle motion and result in complex particle dynamics. Depending on the mode amplitude, as well as on the overlap of nearby resonances, the chaoticity of the particle motion can be either localized or extended, leading to significant modification of the particle distribution function and/or particle losses \citep{Heidbrink2020,White2021}. The origin of the perturbing modes can be either due to externally applied modifications of the magnetic field by using proper sets of coils, or due to intrinsically developed instabilities that may lead to decreased plasma confinement, compromising the high performance operation that is required especially in view of future fusion devices like ITER \citep{Perkins1999}.
Moreover, intentionally introduced perturbative modes can be used in order to control the shape of the distribution function as well as to mitigate undesirable instabilities. For example, Resonant Magnetic Perturbations (RMPs) are widely used to mitigate or suppress pressure-gradient driven Edge Localised Modes (ELMs) \citep{Zohm1996, Liang2007, Evans2008, Wingen2010, Schmitz2012, Orain2019}, inherent to H-mode operation \citep{Wagner1982}, which expunge bursts of particles and heat fluxes towards the Plasma Facing Components (PFC), reducing significantly their lifetime \citep{Loarte2014}.

The effect of non-axisymmetric perturbing modes on particle trajectories, and hence on transport and confinement, is commonly studied on the basis of the Guiding Center (GC) approximation and more specifically on its canonical Hamiltonian formulation \citep{White1982, Littlejohn1983, Meiss1990, Cary2009}. Conditions for resonant interactions strongly depend on the three constants of the motion of the unperturbed GC motion in an axisymmetric magnetic equilibrium, namely the energy $E$, the magnetic moment $\mu$ and the canonical toroidal momentum $P_\zeta$ \citep{Kaufman1972}. Therefore, particles may, or may not, resonantly interact with a mode, depending on their different kinetic characteristics. The resonance conditions for low-energy particles that mostly follow magnetic field lines, can be determined by the helicity of the equilibrium magnetic field, described by the safety factor $q$. However, for more energetic particles that significantly drift from the magnetic field lines, an effective helicity of the drift particle orbit, defined as the kinetic-$q$ factor \citep{Gobbin2008}, is involved in the resonance conditions. The latter is determined by the orbital spectrum of the GC motion \citep{Brizard2014, Zestanakis2016, Antonenas2021} that is the poloidal and bounce/transit-averaged toroidal precession frequencies of trapped and passing particles.

The presence of an electric field has been shown to be closely related to edge magnetic islands introduced by RMPs for ELM mitigation, resulting ambipolar electron/ion transport near islands \citep{Spizzo2014, Ciaccio2015} and being responsible for flows measured during RMP applications \citep{McCool1990, Shaing2002, Evans2015}. Moreover, a drastically increased radial electric field $E_r$ accompanies H-mode operation in fusion plasmas, as was first reported in ASDEX \citep{Wagner1982} and in many other fusion devices ever since. In addition to the strong radial electric field, L-H transition is accompanied by the formation of an Edge Transport Barrier (ETB) across the pedestal of the pressure and density profiles, suppressing the particle transport perpendicularly to the magnetic field and increasing accordingly the plasma pressure and density gradient \citep{Itoh1996}. Due to its efficient confinement characteristics, H-mode operation is considered as the standard baseline scenario for ITER operation \citep{Doyle2007}. Even though a plethora of analytical models have emerged in order to interpret the basic mechanism under L-H transition \citep{Itoh1988, Shaing1989, Connor2000} and many experiments are conducted in order to verify these models \citep{Burrell1997, Connor2000, McDermott2009, Viezzer2013, Viezzer2014, Cavedon2017, Liang2018, Cavedon2020}, the full understanding of the physical mechanism for the enhanced plasma confinement in H-mode operation is an open issue. Theoretical studies have been mainly focused on neoclassical effects following two directions. First, they consider the orbit squeezing effect, according to which the orbit shape of the particle drift can be drastically changed due to the radial electric field so that the ion banana widths can be significantly decreased resulting in reduced prompt losses \citep{Hazeltine1989, Chankin1993, Krasheninnikov1994, Degrassie2009, Brzozowski2019}. Second, the suppression of the perturbation-driven transport due to the formation of ETB caused by the $\mathbf{E\times B}$ shear flow, is investigated. The interplay of the magnetic field shear with the $\mathbf{E\times B}$ shear flow, along with the finite width of the particle drift orbit, determines the kinetic-$q$ factor \citep{Gobbin2008, Zestanakis2016, Ogawa2016, Sanchis2019, Antonenas2021}. The latter dictates the conditions for resonant mode-particle interactions, as well as for the formation of Transport Barriers (TB) at local extrema of the kinetic-$q$ profile, resulting in reduced transport and enhanced confinement \citep{Horton1998, Morrison2000, KurkiSuonio2002, Connor2004, Wagner2006, Putterich2009, Falessi2015, Chen2018, Ida2018, Giannatale2018a, Giannatale2018b, Pegoraro2019}. Thus, kinetic-$q$ factor describes the role of the background magnetic field and the macroscopic flows in plasma transport and stability, under perturbing non-axisymmetric modes and fluctuations. It is worth noting that the dynamics of TB formation is a topic of intense research interest, however most previous works either neglect neoclassical drifts \citep{Rosalem2014, Marcus2019, Grime2023} or consider simplified cylindrical geometries \citep{Ogawa2016}.

The radial electric field is calculated from the lowest order radial force balance equation \citep{Groebner1990, Burrell1997},
\begin{equation}
    E_r=\frac{\nabla P_i}{n_i q_i}-(\mathbf{V}_i\times \mathbf{B})_r=\frac{\nabla P_i}{n_i q_i}-(V_{\theta,i} B_\phi-V_{\phi,i} B_\theta)
\end{equation}
which is valid for any plasma species (or impurity) $i$, with $P, n, q, V_\theta, V_\phi, B_\theta, P_\phi$, being the pressure, density, charge, poloidal and toroidal velocity, poloidal and toroidal magnetic field, corresponding to each species. Experimentally, $E_r$ can be directly measured from the highly localized and accurate signals of Charge eXchange Recombination Spectroscopy (CXRS) diagnostics, with the pressure gradient $\nabla P_i$ being the main contributor to $E_r$ for main ions, and poloidal rotation $V_{\theta,i} B_\phi$ for impurities \citep[\& references therein]{Viezzer2013}. As long as the radial electric field does not depend on the toroidal angle, although it does not perturb the integrability of the GC motion, it strongly modifies the energy landscape of the GC motion. As a consequence, it can significantly modify both the shape of the orbits and their orbital spectrum, and therefore, the kinetic-$q$ factor determining the resonance conditions with non-axisymmetric modes and the overall transport characteristics. 

In this work, we utilize a canonical Hamiltonian GC formulation to systematically study the effect of a radial electric field localized across the pedestal on the particle orbits, under resonant perturbative modes in a toroidal plasma configuration. We investigate the topological changes of the GC phase space through bifurcations of new particle equilibrium points and families of orbits, as well as drastic orbit changes due to the radial electric field that can transform an otherwise lost orbit to a confined one, and vice versa, even in the absence of any perturbation, consequently modifying the particle prompt losses. Moreover, we calculate the orbital spectrum and the kinetic-$q$ factor of the orbits as a function of the constants of the motion for thermal as well as for higher energy particles, taking fully into account the neoclassical drifts, and we show that resonance conditions are drastically altered under the influence of the radial electric field. It is also demonstrated that the radial electric field may prevent or facilitate the resonant mode-particle interactions between specific modes and particles with certain kinetic characteristics. The calculation of the kinetic-$q$ factor enables the accurate prediction of the location of the resonances in the six-dimensional phase space of the system. The a priori knowledge of the resonance locations for a given set of modes is systematically confirmed by numerical particle tracing simulations, suggesting a valuable tool for investigating and designing mode synergies for plasma transport control. Finally, it is shown that the kinetic-$q$ factor can be used for locating the phase space positions where Transport Barriers are formed, persistently bounding the particle orbits and reducing extended particle, energy and momentum transport.    

The paper is organised as follows. In Section \ref{sec:LAR approximation}, the canonical Hamiltonian formulation of the GC motion is presented, while in Section \ref{sec:Radial Electric Field} we describe the configuration of the electric and magnetic field and the respective GC equations of motion. In Section \ref{sec:Change of the GC phase space topology and orbit shape due to Er} we present the modification of the phase space topology and orbit shape due to the radial electric field and its effect on confinement, in the absence of perturbative modes. In Section \ref{sec:Orbital spectrum modification due to Er} we describe the Action-Angle formulation of the GC motion that allows the consistent calculation of the orbital spectrum, the kinetic-$q$ factor and the resonance conditions, and compare the cases of zero and non-zero electric field. In Section \ref{Resonant particle-mode interaction and stochastic transport in the presence of Er} we demonstrate the importance and the accuracy of the a priori knowledge of the resonances and TBs locations, based on the kinetic-$q$ factor, through numerically calculated Poincar{\'e} surfaces of section, under the presence of specific non-axisymmetric modes.

\section{Canonical GC Hamiltonian}\label{sec:LAR approximation}
In a toroidal magnetic configuration consisting of nested magnetic-flux surfaces, the equilibrium magnetic field can be given in contravariant and covariant form using Boozer coordinates \citep{Boozer1981, White1984}
\begin{equation}
    \begin{gathered}
        \mathbf{B}=\nabla\psi\times\nabla\theta+\nabla\zeta\times\nabla\psi_p\\
        \mathbf{B}=g(\psi)\nabla\zeta+I(\psi)\nabla\theta+\delta(\psi,\theta)\nabla\psi
    \end{gathered}
\end{equation}
where $\zeta$ and $\theta$ are the toroidal and poloidal angles, respectively, $\psi$ is the toroidal flux, $\psi_p$ is the poloidal flux, the functions $g(\psi)$, $I(\psi)$ are related to the poloidal and toroidal current, and $\delta(\psi,\theta)$ measures the non-orthogonality of the coordinate system. In such straight-field-line coordinates, the helicity of the field lines is given from the expression
\begin{equation}
    \frac{d\theta}{d\zeta}=\frac{\mathbf{B}\cdot\nabla\theta}{\mathbf{B}\cdot\nabla\zeta}=\frac{1}{q(\psi)}
\end{equation}
with $d\psi_p/d\psi=1/q(\psi)$ and $q(\psi)$ being the safety factor. 

The GC Lagrangian of a charged particle in the presence of electromagnetic fields is given as   $\mathcal{L}=(\mathbf{A}+\rho_\| \mathbf{B})\cdot\mathbf{V}+\mu\dot{\xi}-\mathcal{H}$, where $\mathbf{A}$, $\mathbf{B}$ are the vector potential and the corresponding magnetic field, respectively, $\rho_\|$ is the velocity component parallel to the magnetic field, normalized to $B$, $\mathbf{V}$ is the velocity of the guiding center, $\mu=v_\perp^2/2B$ is the magnetic moment, conjugate to the gyro-phase $\xi$, and
\begin{equation}\label{eq:Hamiltonian_01}
    \mathcal{H}=\rho_\|^2 B^2/2 + \mu B + \Phi
\end{equation}
is the Hamiltonian of the system, with the magnetic field being normalized to its value on magnetic axis $B_0$, and $\Phi$ the electric potential \citep{Littlejohn1983}. The GC motion is expressed in normalized units, namely, time is normalized to inverse cyclotron angular frequency $\omega_0^{-1}$, with $\omega_0=q_i B_0/m_i$ the on axis gyro-frequency, $q_i$ and $m_i$ the ion charge and mass, respectively, lengths are normalized to the major axis $R_0$, and consequently, energy is normalized to $m_i\omega_0^2R_0^2$. 

The canonical momenta conjugate to angles $\zeta$ and $\theta$, are properly expressed in relation to the variables $\rho_\|$, $g(\psi)$ and $I(\psi)$ \citep[p. 79]{White2014}, as
\begin{align}
    P_\zeta &= g(\psi)\rho_\|-\psi_p(\psi)\label{eq:P_zeta} \\
    P_\theta &= \psi+\rho_\|I(\psi)\label{eq:P_theta} 
\end{align}
and consequently, the Hamiltonian \eqref{eq:Hamiltonian_01} is given in the form
\begin{equation}\label{eq:Hamiltonian_02}
    \mathcal{H} = \frac{(P_\zeta+\psi_p(P_\zeta,P_\theta))^2}{2g^2(P_\zeta,P_\theta)} B^2(P_\zeta,P_\theta,\zeta, \theta) + \mu B(P_\zeta,P_\theta,\zeta,\theta) + \Phi(P_\zeta,P_\theta,\zeta,\theta)
\end{equation}
The absence of an explicit time dependence renders the Hamiltonian, and hence the energy $E$ of the system, a constant of the motion, whereas the absence of the gyro-phase $\xi$ results in the conservation of the magnetic moment $\mu$. In the case of an axisymmetric configuration where both the magnetic field and the electric potential are independent of the toroidal angle $\zeta$, the  conjugate canonical momentum $P_\zeta$ is an additional invariant of the motion. The existence of three independent constants of the motion $(E,\mu,P_\zeta)$ renders the system integrable and along with the sign of $\rho_\|$, are uniquely defining, hence labelling, the orbits of the particles in phase space. 

In the following sections, our calculations are carried out under the consideration of a circular cross-section Large Aspect Ratio (LAR) magnetic field equilibrium, with $\epsilon=r_0<<1$ being the inverse aspect ratio and $r_0$ the minor axis (normalized to major axis $R_0$). Under this approximation, it is easy to show that, to leading order, $I(\psi)\simeq0$, $g(\psi)\simeq1$, $\delta(\psi,\theta)=0$ \citep{White1984},\citep{White2013b}. Subsequently, from Eq. \eqref{eq:P_theta} we get $P_\theta=\psi=r^2/2$, while the magnetic field acquires the quite simple form $B=B_0(1-r\cos\theta)=B_0(1-\sqrt{2P_\theta}\cos\theta)$.

\section{Radial electric field across the pedestal}\label{sec:Radial Electric Field}
We consider a strong,  radially dependent and highly localized non-monotonic electric field, neglecting any dependence on the poloidal or toroidal angle, which is given from the expression
\begin{equation}\label{eq:Er}
    E_r(r)=-E_a\exp\left[-\frac{(r-r_a)^2}{r_w^2}\right],
\end{equation}
where $E_r(r)$ is the well-like profile of the electric field, $E_a$ is the amplitude (depth) of the well located at $r=r_a$, and $r_w$ is the waist of the radial profile. Adopting typical values for the electric field profile near the wall \citep{KurkiSuonio2002, Burrell2004, McDermott2009, Viezzer2013, Huang2020} and disregarding its shape close to the core, the respective well is placed at $r_a=0.98\,r_0$, having a waist $r_w=r_0/50$ and an amplitude $E_a\simeq 75\un{kV/m}$, deliberately selected to yield an electrostatic potential of approximately $\Phi= 1.2\un{kV}$ at the Last Closed Flux Surface (LCFS) denoted as $\psi_{wall}$. 

\begin{figure}
    \centering
    \includegraphics[width=0.90\textwidth]{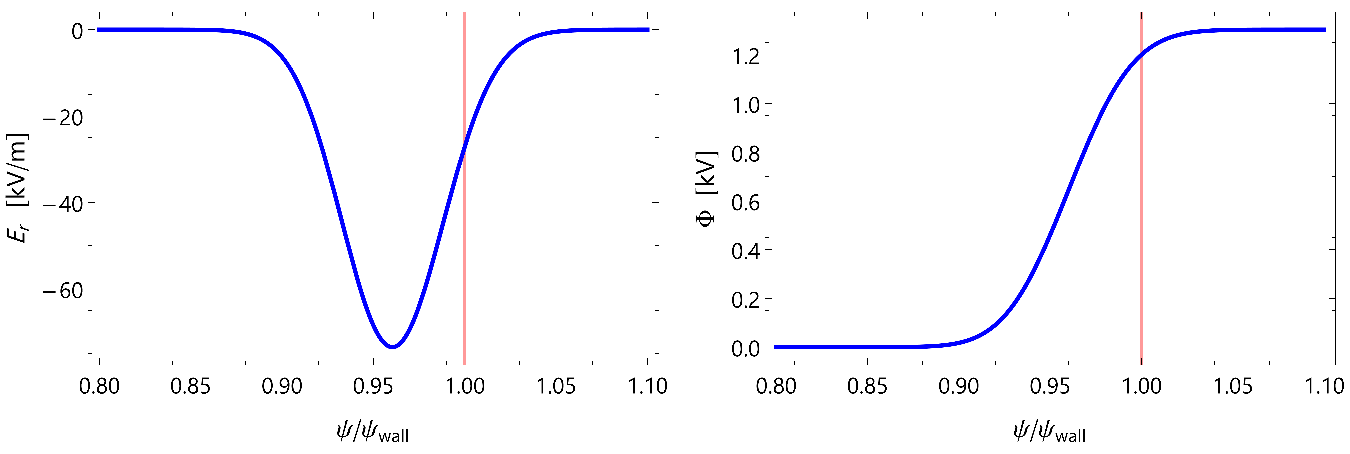}
    \caption{Radial profiles of the radial electric field $E_r(\psi)$ (left) and of the respective electric potential $\Phi(\psi)$ (right). For a LAR magnetic field equilibrium $\psi=P_\theta$. The red vertical line marks the wall.}
    \label{fig:Er_Phir}
\end{figure}

The electric potential corresponding to Eq. \eqref{eq:Er} is straightforwardly calculated from the integral
\begin{equation}
    \Phi(r)=-\int_0^r E_r(r')\,dr',
\end{equation}
which in terms of the poloidal momentum $P_\theta=\psi=r^2/2$ (for a LAR magnetic field equilibrium) it yields
\begin{equation}\label{eq:Electric_potential}
    \Phi(P_\theta)=E_a\sqrt{\frac{\pi P_{\theta w}}{2}}\left[\mathrm{erf}\left(\frac{\sqrt{P_\theta}-\sqrt{P_{\theta a}}}{\sqrt{P_{\theta w}}}\right) + \mathrm{erf}\left(\sqrt\frac{{P_{\theta a}}}{P_{\theta w}}\right)\right] 
\end{equation}
with $P_{\theta w}=r_w^2/2$ and $P_{\theta a}=r_a^2/2$ being the waist and the location of the well, respectively. Figure~\ref{fig:Er_Phir} shows the radial profile of the radial electric field (left) as well as the respective potential barrier (right), both as function of the toroidal flux $\psi$, coinciding with the canonical poloidal momentum $P_\theta$ for a LAR equilibrium.

Under the aforementioned approximations of a LAR equilibrium magnetic field and a purely radial electric field, the GC Hamiltonian takes the form
\begin{align}\label{eq:Hamiltonian_LAR}
    \mathcal{H} = \frac{\left(P_\zeta+\psi_p(P_\theta)\right)^2}{2} (1-\sqrt{2P_\theta}\cos\theta)^2 + \mu(1-\sqrt{2P_\theta}\cos\theta) + \Phi(P_\theta),
\end{align}
where the axisymmetry of the dynamical system is preserved due to the absence of the canonical angle $\zeta$. The corresponding equations of motion are derived from the expressions
\begin{align}
    \dot{P}_\zeta&=-\frac{\partial\mathcal{H}}{\partial\zeta} = 0\label{eq:dPzeta|dt}\\    
    \dot{P}_\theta&=-\frac{\partial\mathcal{H}}{\partial\theta} = -(P_\zeta+\psi_p(P_\theta))^2(1-\sqrt{2P_\theta}\cos\theta)\sqrt{2P_\theta}\sin\theta -\mu\sqrt{2P_\theta}\sin\theta \\
    \dot{\zeta}&=\frac{\partial\mathcal{H}}{\partial P_\zeta} =(P_\zeta+\psi_p(P_\theta))(1-\sqrt{2P_\theta}\cos\theta)^2 \label{eq:dzeta|dt}\\
    \dot{\theta}&=\frac{\partial\mathcal{H}}{\partial P_\theta} = -\frac{(1-\sqrt{2P_\theta}\cos\theta)(P_\zeta+\psi_p(P_\theta))^2}{\sqrt{2P_\theta}} -\frac{\mu\cos\theta}{\sqrt{2P_\theta}}+\notag\\
    &\qquad\qquad\qquad\qquad\qquad +(1-\sqrt{2P_\theta}\cos\theta)^2(P_\zeta+\psi_p(P_\theta))\psi_p'(P_\theta) + \Phi'(P_\theta)
    \label{eq:dtheta|dt}
\end{align}
where primes imply differentiation with respect to $P_\theta$. Equation \eqref{eq:dtheta|dt} clearly shows that the radial electric field effects are introduced exclusively through the $\Phi'(P_\theta)$ term of the poloidal angle time-derivative $\dot\theta$, appearing additively to the magnetic shear effects, as expressed by the term proportional to the $\psi'_p(P_\theta)$, a fact that has a significant impact on the modification of GC orbits in terms of their shape and poloidal angular frequency, as will be shown in the following sections. 

The magnetic shear is dictated by the $q$ factor profile, which in our analysis is considered according to the commonly used expression \citep[p.56]{White2014} 
\begin{equation}
    q(P_\theta)=q_0\left[ 1+\left(\frac{P_\theta}{P_{\theta k}(q_{wall})}\right)^n \right]^{1/n},
\end{equation}
where $q_0$ is the value of the safety factor on the magnetic axis, $P_{\theta k}$ is a function related to the location of the "knee" of the profile, and $n$ characterizes the order of the equilibrium, with $n=1,2,4$ denoting a peaked, round, or flat profile, respectively. In that respect, the poloidal flux can be analytically obtained by integrating the rotational transform $i(P_\theta)=1/q(P_\theta)$
\begin{equation}
    \psi_p(P_\theta)=\int i(P_\theta) dP_\theta
\end{equation}
which results in the hypergeometric function ${}_2F_1$ of the general form 
\begin{equation}\label{eq:psi_p}
    \psi_p(P_\theta)=\frac{P_\theta}{q_0}\,{}_{2}F_1\left[\frac{1}{n},\frac{1}{n};1+\frac{1}{n},\left(1-\left(\frac{q_{wall}}{q_0}\right)^n\right)\left(\frac{P_\theta}{P_{\theta wall}}    \right)^n\right],
\end{equation}
where $q_{wall}$ and $P_{\theta wall}$ are the values of the safety factor and the poloidal momentum on the wall, respectively. In the following analysis we select a $q$ profile with the moderate $n=2$ value, $q_0=1.1$, and $q_{wall}=3.5$.

The consideration of the LAR equilibrium, along with the corresponding simplification $\psi=P_\theta$ and the analytical expression \eqref{eq:psi_p} for $\psi_p(P_\theta)$, result in an explicit expression of the equations of motion. It is important to note that for a generic (non-LAR) equilibrium, the Hamiltonian cannot be explicitly expressed in canonical variables, instead, one should use the equations of motion in terms of $(\theta, \psi_p, \rho_\parallel, \zeta)$ \citep{White2014} and afterwards calculate the canonical momenta $P_\theta$, $P_\zeta$, according to \eqref{eq:P_zeta}, \eqref{eq:P_theta}. In order to solve the equations of motion \eqref{eq:dPzeta|dt}-\eqref{eq:dtheta|dt}, a fourth-order Runge-Kutta method is implemented, also used in well-known orbit following codes like ORBIT \citep{White1984} or ASCOT \citep{Hirvijoki2014}.

\section{Change of the GC phase space topology and orbit shape due to $E_r$}\label{sec:Change of the GC phase space topology and orbit shape due to Er}

The existence of a radial electric field that is localized near the LCFS of the tokamak, affects the topology of the GC phase space by introducing additional fixed points and by modifying the shape of the GC orbits. As a result, the presence of the electric field can significantly modify the prompt particle losses by changing an otherwise confined orbit into a lost one, and vice versa \citep{Hazeltine1989, Chankin1993, Krasheninnikov1994}. 

For typical values of $E_r$ amplitude ($10-60\un{kV/m}$) and waist ($1-2\un{cm}$, restricted in ranges $0.85 r_0-1.05 r_0$), the corresponding electric potential energy is of the order of a few hundreds of $\un{eV}$. Even though these values of the potential energy are not sufficient to essentially reshape the highly energetic particle trajectories, the impact to thermal and mildly energetic particles, located well inside the bulk of the distribution function profile, could considerably affect the confinement inside this region. Apparently, this impact depends strongly on the amplitude of the potential barrier and will increase as it becomes comparable to the kinetic energy of the particles, as well as on the effective time that a particle spends inside the well during its radial drift. 

\begin{figure}
     \centering
     \begin{subfigure}[t]{0.90\textwidth}
         \centering
         \includegraphics[width=\textwidth]{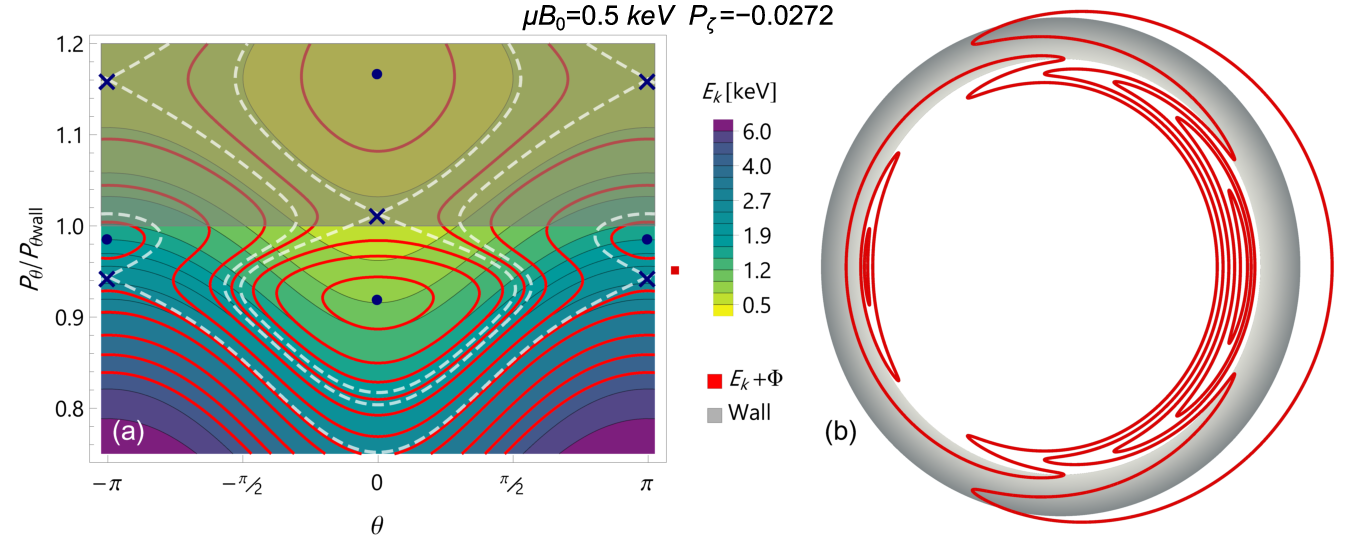}
     \end{subfigure}     
     \begin{subfigure}[t]{0.90\textwidth}
         \centering
         \includegraphics[width=\textwidth]{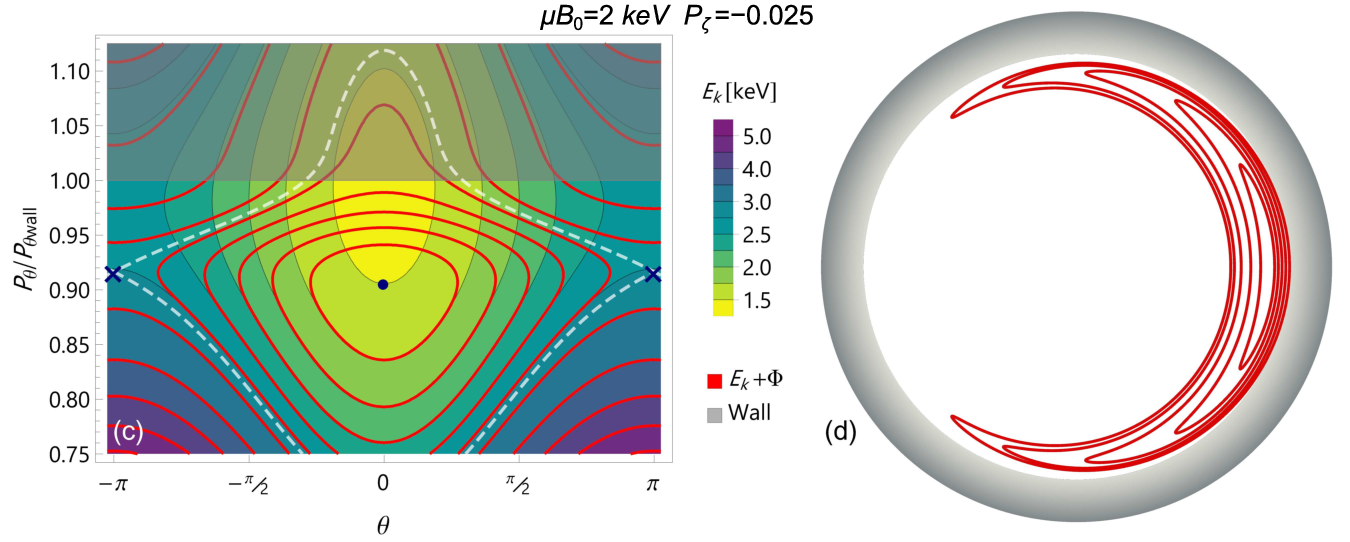}
     \end{subfigure}     
     \begin{subfigure}[t]{0.90\textwidth}
         \centering
         \includegraphics[width=\textwidth]{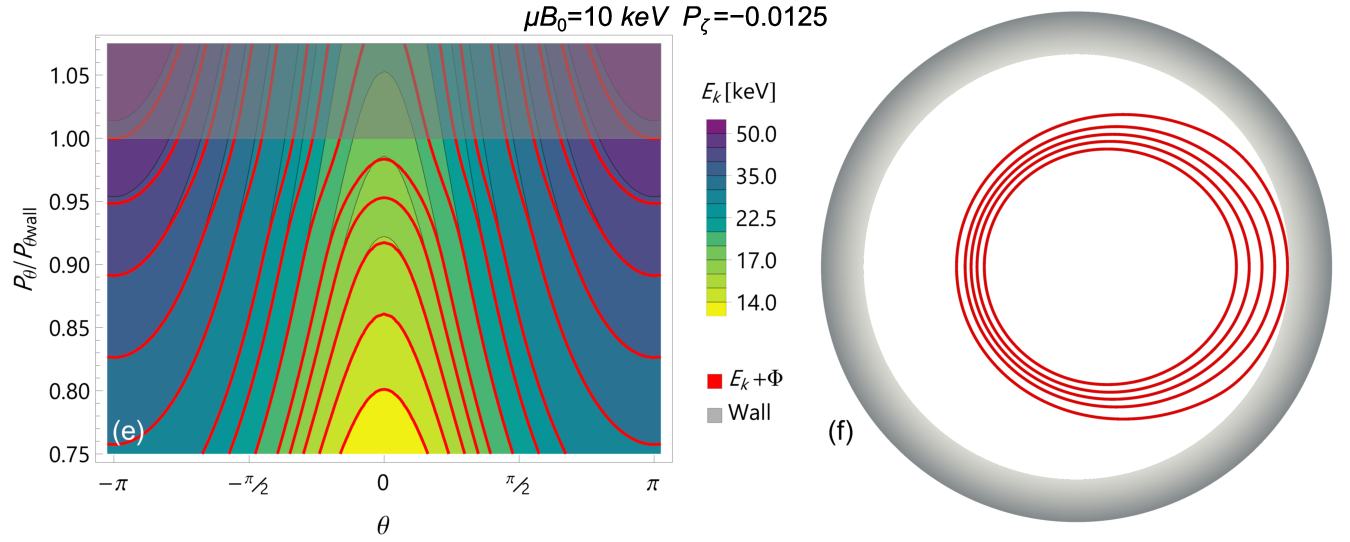}
     \end{subfigure}
        \caption{(a, c, e) GC orbits depicted in phase space, with background coloured orbits corresponding to $E_r=0$ and red orbits corresponding to $E_r\neq0$. (b, d, f) Characteristic GC orbits for $E_r\neq 0$, depicted in configuration space. Dots and X's denote elliptic and hyperbolic critical points, respectively, and white dashed lines denote the trapped-passing boundaries (separatrices) when $E_r\neq0$. 
        (a, b) Thermal particles with $\mu B_0=0.5\un{keV}$, $P_\zeta=-0.0272$. $E_r$ causes the emergence of additional critical points near the wall, inducing 
        additional trapped orbits along $\theta=\upi$ midplane and twice-reversed passing particles manifesting the abrupt changes of the perpendicular drift velocity. (c, d) Low-energy particles with $\mu B_0=2\un{keV}$, $P_\zeta=-0.025$. $E_r$ rearranges the orbits into the confined trapped domain. (e, f): Energetic particles with $\mu B_0=10\un{keV}$, $P_\zeta=-0.0125$. As the kinetic energy increases, the contribution of the potential to the total energy decreases, leaving the phase space of such orbits practically undisturbed. The confined passing orbits retain their shapes, and only minor modifications near the wall witness the presence of $E_r$.}
        \label{fig:phaseSpaceConfigurationSpace}
\end{figure}

When no electric potential exists, the $\theta-P_\theta$ plane of the phase space of the LAR Hamiltonian demonstrates a single elliptical point along the outboard midplane $\theta=0$, with its location along $P_{\theta}$ depending on $\mu$ and $P_\zeta$. Evidently, the existence of a radial electric field can effectively modify this picture in two ways. First, the inclusion of the potential term $\Phi(P_\theta)$ in the Hamiltonian may impose significant changes in the topology of the phase space, in the sense that either the location of the respective critical points may be rearranged, or additional critical points may emerge due to bifurcations. In either case, the transformation of the phase space could significantly affect the respective orbits in terms of their characterization as passing or trapped, and confined or lost. Second, when the elliptic point of the system lies in the neighborhood of the pedestal, the lower-energy orbits residing relatively close to the elliptic point will undergo the strongest impact, as their energy will be comparable to the amplitude of the potential barrier. Moreover, due to their relatively small radial drift, these particles will remain well inside the vicinity of the strong potential barrier along their entire orbits, increasing the effective interaction time of the system. On the other hand, higher energy particles undergo a wider radial drift motion inside the torus, and therefore their trajectories spend a smaller fraction of their period inside the pedestal, decreasing the influence that $E_r$ has on the respective orbits.

These features are very clearly illustrated in Fig. \ref{fig:phaseSpaceConfigurationSpace}, where in the left column we show the $\theta-P_\theta$ projection of the phase space of the Hamiltonian \eqref{eq:Hamiltonian_LAR}, for various values of the perpendicular kinetic energy $\mu B_0$ and the toroidal momentum $P_\zeta$, with the latter being properly selected in order for the elliptical point to be located in the vicinity of the potential barrier, where its effects are more acute. Coloured contour-plots correspond to the kinetic energy $E_k$ levels of Eq. \eqref{eq:Hamiltonian_02}, when $\Phi(P_\theta)=0$, and red contour-lines correspond to energy levels with $\Phi(P_\theta)$ being given from Eq. \eqref{eq:Electric_potential}. Accordingly, in the right column, the respective orbits are depicted in the configuration space of the LAR magnetic equilibrium. Grayed areas denote the wall (LCFS)  of the equilibrium, and no re-entrance is allowed for particles beyond this point as they are considered to be lost. As it is clearly seen, the modification of the phase space trajectories is more vividly manifested for thermal and weakly energetic particles initialized along the potential gradient, whereas higher energy particles experience only minor modifications in their orbits, in the form of small variations from the original $E_r=0$ case.

Focusing on the implications that the radial electric field imposes upon the particles orbits for various sets of parameters, in the first row of Fig. \ref{fig:phaseSpaceConfigurationSpace} we depict the orbits of thermal particles with $\mu B_0=0.5\un{keV}$,  $P_\zeta=-0.272$, for various values of the total energy $E$. In panel (a) it is shown that in the absence of $E_r$ (coloured orbits), the trapped/passing boundary is formed beyond the wall, hence, there exist low-energy counter-passing orbits being confined for small $P_\theta$ and eventually lost as $P_\theta$ increases. 
In the presence of $E_r$ (red orbits), the picture changes significantly, and confined trapped orbits populate the phase space near the wall, as a result of the emergence of an additional elliptical point inside of the wall, along outboard midplane $\theta=0$. Moreover, the appearance of a pair of critical points close to the wall, an elliptic along $\theta=\pm\upi$, and  a hyperbolic along $\theta=0$, respectively, is responsible for the appearance of a new family of orbits oscillating around the elliptic point at $\theta=\pm\upi$, as well as for the unusual parallel velocity reversal points encountered by passing orbits, manifested also as turning points in the poloidal direction of the configuration phase [panel (b)].

The existence of turning points in the pedestal, is associated with the contribution of $\mathbf{E}\times\mathbf{B}$ drift velocity to the total perpendicular velocity of the plasma particles and the conservation of the magnetic moment $\mu$. Normally, for a passing orbit in a torus, the initial velocity ratio $v_{\parallel 0}/v_{\perp 0}$ is sufficiently large to prevent the existence of a turning point, however, when a radial electric field is present, the sudden increase of the perpendicular drift velocity $V_{E\times B}$ inside the pedestal, results in the decrease of the velocity ratio, allowing now the velocity reversal due to $\nabla B$ drift along the orbit. As soon as the influence of the electric field starts to decline, the particle goes through its second turning point, carrying out a full rotation around the magnetic axis.

In the second row of Fig. \ref{fig:phaseSpaceConfigurationSpace}, we depict low-energy particles with $\mu B_0=2\un{keV}$ and $P_\zeta=-0.025$. In the absence of $E_r$ (coloured orbits), the dominant elliptic point roughly touches the wall, and consequently, the near-wall area is populated by either confined counter-passing or promptly-lost trapped particles. In the presence of the radial electric field (red orbits), the existence of the elliptic point at $\theta=0$, $P_\theta\simeq 0.9 P_{\theta,wall}$, drastically changes the topology of the phase space, allowing for the banana orbit width squeezing \citep{Hazeltine1989, Chankin1993, Krasheninnikov1994} and the existence of confined trapped orbits, as well.

Finally, the third row depicts orbits of higher-energy particles with $\mu B_0=10\un{keV}$ and $P_\zeta=-0.0125$. In this instance, the kinetic energy of the particles dominates over the potential barrier, and counter-passing orbits populate almost exclusively the outer area of the torus, regardless of the presence of the radial electric field. The relatively large radial drift undergone by the particles allows only a partial interaction with the potential gradient, which is constrained within the narrow pedestal region. This weak interaction is graphically witnessed in both the phase space (e) and the configuration space (f), by observing the minor deviations between the respective orbits in the absence (coloured orbits), and in the presence (red orbits) of $E_r$.

\section{Orbital spectrum modification due to $E_r$}\label{sec:Orbital spectrum modification due to Er}
The electric potential term of the Hamiltonian does not affect the integrability property of the GC motion, however, as shown in Eq. \eqref{eq:dtheta|dt}, it directly affects the time evolution of the poloidal angle and modifies the poloidal period, as well as the periods of the other degrees of freedom that are coupled to the poloidal motion. Hence, apart from the shape variations that a trajectory is undergone due to the existence of the potential barrier, the frequency spectrum of the respective orbits may significantly deviate from the one obtained in the absence of $E_r$, and consequently, rearrange dramatically the resonance condition of the particle orbits with non-axisymmetric perturbations. This aspect is of paramount importance as the resonant mode-particle interactions dictate momentum and particle transport phenomena in fusion plasmas \citep{Heidbrink2020, White2021}.

In the following, we investigate the impact of the radial electric field on the orbital spectrum, namely the poloidal and the bounce/transit-averaged toroidal precession frequencies, of different orbits, characterized by a specific set of constants of the motion ($\mu, E, P_\zeta$), and therefore, the respective drift-mode resonant numbers corresponding to these orbits. To perform our calculations, we take advantage of the Hamiltonian framework by describing our system in the Action-Angle variables formalism which fully exploits the canonical structure of GC dynamics \citep{Zestanakis2016, Antonenas2021}.

The integrability of the GC Hamiltonian \eqref{eq:Hamiltonian_LAR}, allows for a canonical transformation from the original canonical variables to Action-Angle variables
\begin{equation}\label{eq:Canon2ActionVars}
    \begin{gathered}
        (\mu, \xi)\rightarrow(J_\xi, \hat{\xi}) \\
        (P_\zeta, \zeta)\rightarrow(J_\zeta, \hat{\zeta}) \\
        (P_\theta, \theta)\rightarrow(J_\theta, \hat{\theta}) \\
    \end{gathered}
\end{equation}
using the appropriate mixed-variable generating function \citep{Lichtenberg1992, Goldstein2002} 
\begin{equation}
    F_2(J_\xi, J_\zeta, J_\theta, \xi, \zeta, \theta) = \xi J_\xi + \zeta J_\zeta + f_2(J_\xi, J_\zeta, J_\theta, \theta),
\end{equation}
with
\begin{equation}
    f_2(J_\xi, J_\zeta, J_\theta, \theta) = f_2(\mathbf{J,\theta})=\int^{\theta}P_\theta\bigl(J_\xi, J_\zeta, J_\theta(J_\xi, J_\zeta),\theta'\bigr)\,d\theta'.
\end{equation}
It is quite straightforward to show that in the first two sets of Eq. \eqref{eq:Canon2ActionVars}, the respective Actions are equal to the old momenta, that is $\mu=J_\xi$ and $P_\zeta=J_\zeta$, while the third Action is defined as
\begin{equation}\label{eq:J_theta}
    J_\theta=\frac{1}{2\pi}\oint P_\theta(E, J_\xi, J_\zeta,\theta)\,d\theta,
\end{equation}
where $E$ is the total energy of the dynamical system, and $J_\theta$ may correspond to either bounced (trapped) or transit (passing) motion.
    The new angles (canonical positions) corresponding to bounce-averaged gyro-angle, toroidal and poloidal angles, respectively, are given in relation to the old angles as \citep{Kaufman1972, Zestanakis2016}
\begin{equation}\label{eq:Angles}
    \hat{\xi}=\xi+\frac{\partial f_2(\mathbf{J,\theta})}{\partial J_\xi}, \qquad \hat{\zeta}=\zeta+\frac{\partial f_2(\mathbf{J,\theta})}{\partial J_\zeta}, \qquad \hat{\theta}=\frac{\partial f_2(\mathbf{J,\theta})}{\partial J_\theta}.
\end{equation}
It is worth noting that the angles $(\xi, \zeta, \theta)$ do not exhibit a linear time dependence, and hence, the respective frequencies $(\dot{\xi}, \dot{\zeta}, \dot{\theta})$ do not represent any physical frequency of the Hamiltonian system. However, Eq. \eqref{eq:Angles} implies that $\hat{\theta}$ is a periodic function of $\theta$, and as such, both share the same periodicity, yielding
\begin{equation}
    T_{\hat\theta}=T_\theta=2\pi/\hat\omega_\theta, 
\end{equation}
a feature that does not apply to the rest of the angles $\hat{\xi}, \hat{\zeta}$.

Omitting the variables related to fast gyromotion for brevity, the orbital frequencies of the remaining degrees of freedom are acquired by virtue of implicit differentiation 
\begin{align}\label{omega_zeta/omega_theta}
    \frac{\hat{\omega}_\zeta}{\hat{\omega}_\theta}=\frac{\partial\mathcal{H}\bigl(J_\xi,J_\zeta,J_\theta(J_\xi,J_\zeta)\bigr)/\partial J_\zeta}{\partial\mathcal{H}\bigl(J_\xi,J_\zeta,J_\theta(J_\xi,J_\zeta)\bigr)/\partial J_\theta} = -\frac{\partial J_\theta(J_\xi, J_\zeta)}{\partial J_\zeta}
\end{align}
revealing the fact that under the Action-Angle formalism the poloidal Action $(-J_\theta)$ can be considered as the new Hamiltonian of the system \citep{White1984, Antonenas2021}, with the time being normalized to the inverse poloidal frequency $\hat\omega_\theta$. 
Substituting Eq. \eqref{eq:J_theta} into Eq. \eqref{omega_zeta/omega_theta} yields
\begin{align}\label{eq:omega_zeta/omega_theta}
    \frac{\hat{\omega}_\zeta}{\hat{\omega}_\theta} &= -\frac{1}{2\upi}\frac{\partial}{\partial J_\zeta}\oint P_\theta(E, \mu, P_\zeta, \theta)\,d\theta 
    = -\frac{1}{2\upi}\oint \frac{\partial P_\theta(E, \mu, P_\zeta, \theta)}{\partial J_\zeta}\,d\theta \nonumber\\ 
    &=\frac{1}{2\upi}\oint\frac{1}{\dot{\theta}}\frac{\partial\mathcal{H}}{\partial P_\zeta}\,d\theta=\frac{1}{2\upi}\int_0^{T_\theta}\dot{\zeta}\,dt = \frac{1}{2\upi}\int_{\zeta_1(t=0)}^{\zeta_2(t=T_\theta)}\,d\zeta, \nonumber\\
     \frac{\hat{\omega}_\zeta}{\hat{\omega}_\theta} &=\frac{(\Delta\zeta)_{T_{\theta}}}{2\upi}
\end{align}
Considering also that
\begin{equation}\label{eq:Averaged_dzeta|dt}
    \langle\dot{\zeta}\rangle_{T_\theta} \doteq \frac{1}{T_\theta}\int_0^{T_\theta}\dot{\zeta}\,dt = \frac{1}{T_\theta}\int_{\zeta_1(t=0)}^{\zeta_2(t=T_\theta)}\,d\zeta =\frac{\hat\omega_\theta}{2\pi}(\Delta\zeta)_{T_\theta},
\end{equation}
Eqs. \eqref{eq:omega_zeta/omega_theta} and \eqref{eq:Averaged_dzeta|dt} lead us to the conclusion that 
\begin{equation}
    \hat\omega_\zeta=\dot{\hat\zeta}=\langle\dot\zeta\rangle_{T_\theta},
\end{equation}
suggesting that the frequency $\hat\omega_\zeta$ is directly associated with the bounce/transit averaged toroidal precession \citep{White2015,Antonenas2021}.

The toroidal to poloidal frequency ratio calculated in Eq. \eqref{eq:omega_zeta/omega_theta} is associated with the kinetic resonance condition of the unperturbed GC particle motion with any non-axisymmetric perturbative mode of the form $F_{mn}(\psi)\exp[i(n\hat{\zeta}-m\hat{\theta})]$  
\begin{align}\label{eq:q_kin}
    &n\hat{\omega}_\zeta-m\hat{\omega}_\theta=0,\nonumber\\
    &q_{kin}\equiv\frac{\hat{\omega}_\zeta}{\hat{\omega}_\theta}=\frac{m}{n},
\end{align}
with $q_{kin}$ being the drift-frequency ratio, also known as the kinetic-$q$ factor, and $m,n$ being the (integer) poloidal and toroidal mode numbers, respectively. The concept of the kinetic-$q$ factor, first introduced by \citep{Gobbin2008}, has a paramount role in resonant mode-particle interactions \citep{White2014}, and its significance was first reported when a discrepancy was detected between the location of the resonant orbits and the location of the resonant surfaces of the magnetic $q$ profile, in MST reversed field pinch \citep{Fiksel2005}. The difference between kinetic and magnetic $q$ factors implies that kinetic and magnetic resonant islands, along with the accompanied chaoticity, may occur in different spatial positions, depending on the kinetic characteristic of each orbit, namely the energy, the magnetic moment, and the poloidal and toroidal momenta. In particular, higher energy particles experience significant radial drifts across the magnetic field lines, and their trajectory only partially resides in the vicinity of a magnetic island, suggesting that the chaoticity of the orbits cannot be directly derived from the chaoticity of the magnetic field lines. This is not the case for low-energy particles, whose radial drift is considerably smaller, and hence, particle trajectories follow the magnetic field lines, inheriting its chaoticity  \citep{Cambon2014, He2019, He2020, Moges2024}.

\begin{figure}
    \centering
    \includegraphics[width=0.90\textwidth]{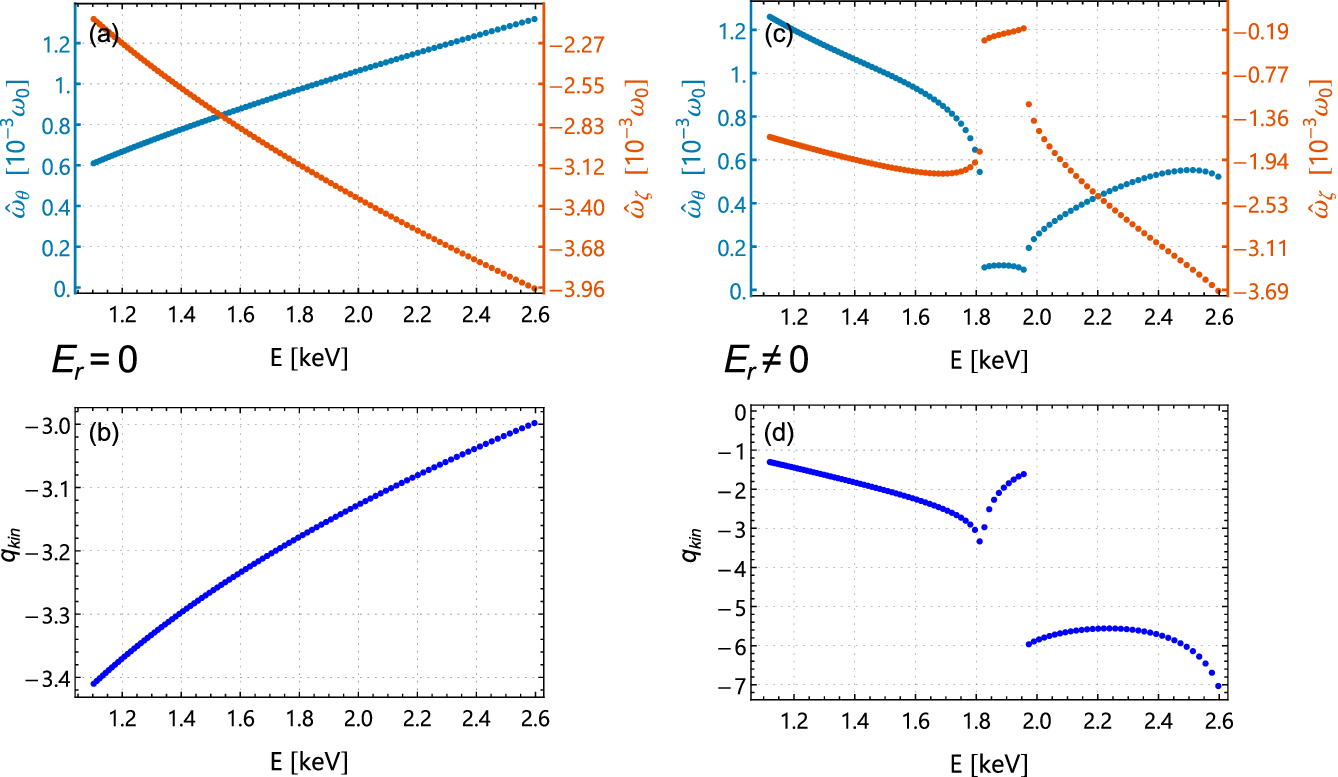}
    \caption{Effects of $E_r$ on the orbital spectrum and the kinetic-$q$ factor ($q_{kin}$) for thermal particles, with $\mu B_0=0.5\un{keV}$, $P_\zeta=-0.0272$, corresponding to figure \ref{fig:phaseSpaceConfigurationSpace}(a,b). 
    (a, c) Poloidal frequency $\hat{\omega}_\theta$ (light blue points) and toroidal precession frequency $\hat{\omega}_\zeta$ (orange points), as functions of the energy $E$, for $E_r=0$ (a) and $E_r \neq 0$ (c). 
    (b, d) Kinetic-$q$ factor $q_{kin}=m/n=\hat{\omega}_\zeta/\hat{\omega}_\theta$, as a function of the energy $E$, for $E_r=0$ (b) and $E_r \neq 0$ (d). Multiple branches correspond to different orbit families.}
    \label{fig:SpectrumResonances__muB0_0.5_keV}
\end{figure}

\begin{figure}
    \centering
    \includegraphics[width=0.90\textwidth]{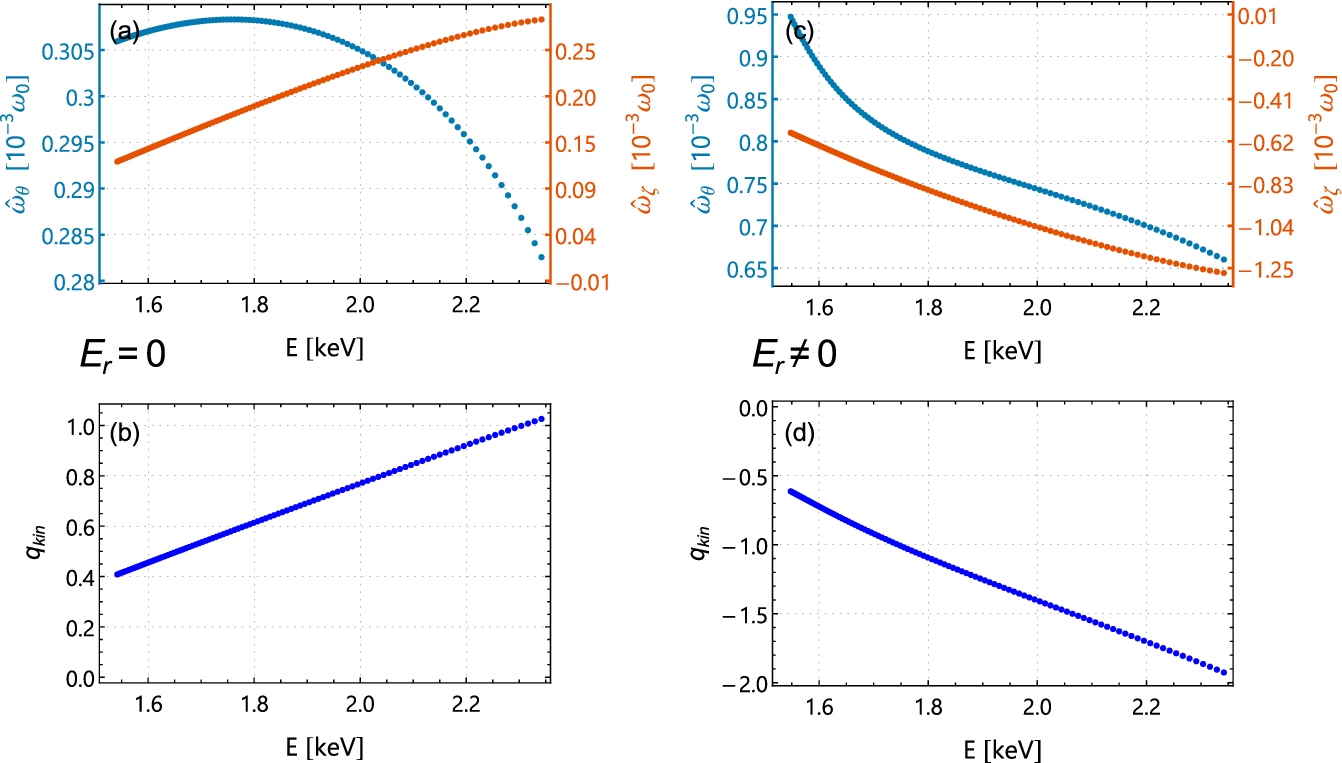}
    \caption{Same as figure \ref{fig:SpectrumResonances__muB0_0.5_keV} for low energy particles with $\mu B_0=2\un{keV}$, $P_\zeta=-0.025$, corresponding to figure \ref{fig:phaseSpaceConfigurationSpace}(c, d). The presence of the radial electric field changes the sign and the monotonicity of the $q_{kin}(E)$ curve.}
    \label{fig:SpectrumResonances__muB0_2.0_keV}
\end{figure}

\begin{figure}
    \centering
    \includegraphics[width=0.90\textwidth]{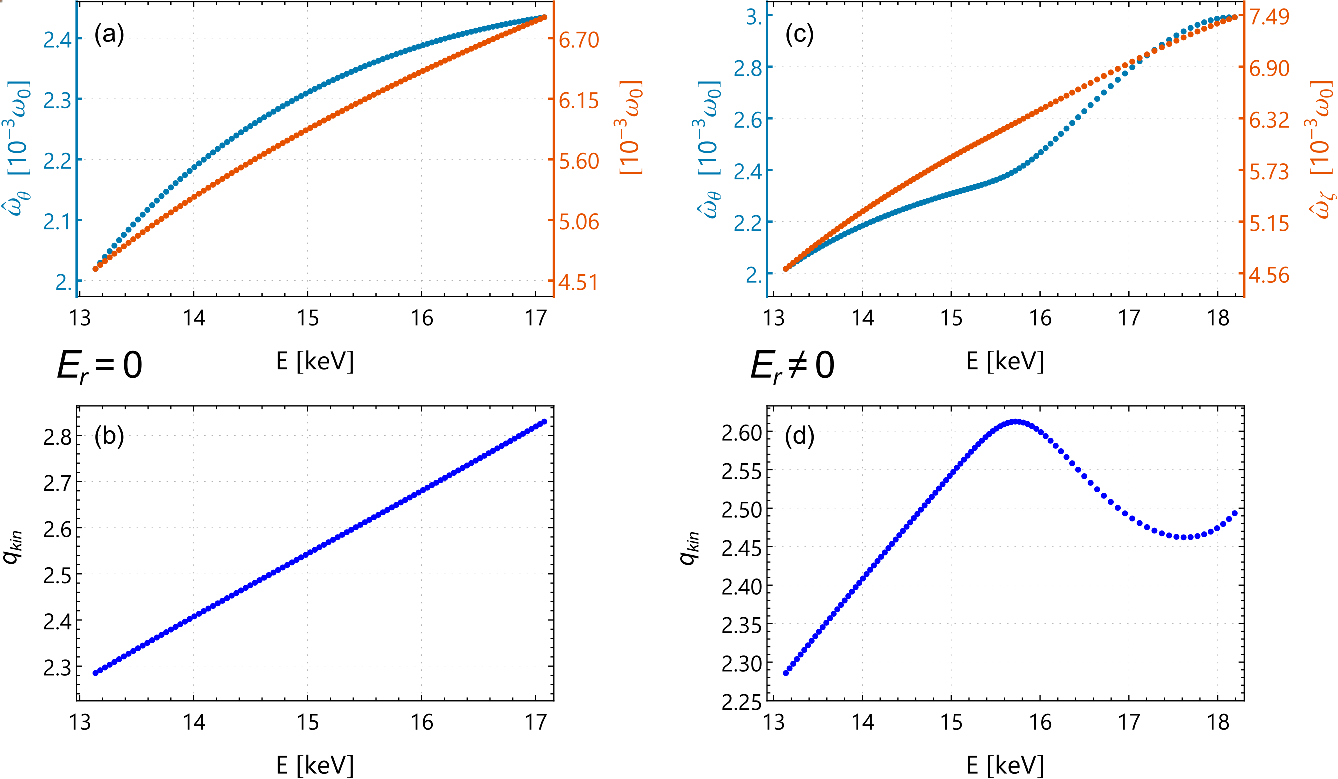}
    \caption{Same as figure \ref{fig:SpectrumResonances__muB0_0.5_keV} for mildly energetic particles with $\mu B_0=10\un{keV}$, $P_\zeta=-0.0125$ corresponding to figure \ref{fig:phaseSpaceConfigurationSpace}(e, f). The presence of the radial electric field renders the $q_{kin}(E)$ non-monotonic and introduces local extrema with important implications for stochastic transport related to Stochastic Transport Barriers.}
    \label{fig:SpectrumResonances__muB0_10.0_keV}
\end{figure}

In order to corroborate our arguments, regarding the expected changes of the orbital spectrum and the kinetic-$q$ factor, in this section we present indicative cases, corresponding to those shown in Fig. \ref{fig:phaseSpaceConfigurationSpace}. In the left column of Figs. \ref{fig:SpectrumResonances__muB0_0.5_keV}-\ref{fig:SpectrumResonances__muB0_10.0_keV} we present the orbital spectrum, that is the  poloidal and bounce/transit-averaged toroidal precession frequencies in panels (a), as well as the kinetic-$q$ factor $(q_{kin})$ in panels (b), in the absence of the radial electric field. Accordingly, right column depicts the same information, in panels (c) and (d), when a radial electric field is present.

In particular, Fig. \ref{fig:SpectrumResonances__muB0_0.5_keV} depicts the orbital frequencies and the kinetic-$q$ factor for thermal particles corresponding to Figs. \ref{fig:phaseSpaceConfigurationSpace}(a),(b), with $\mu B_0=0.5\un{keV}$ and $P_\zeta=-0.0272$, in the absence [(a),(b)] and in the presence [(c),(d)] of the radial electric field. The existence of a non-zero $E_r$ changes both quantitatively and qualitatively the orbital frequencies. The drastic qualitative changes, manifested by multiple branches, correspond to the introduction of additional orbit families due to the radial electric field. Kinetic-$q$ is modified accordingly, suggesting that the radial electric field drastically changes the conditions for resonant particle interaction with non-axisymmetric modes, since Eq. \eqref{eq:q_kin} can now be fulfilled for different mode numbers $(m,n)$ and particle energies. According to Fig. \ref{fig:SpectrumResonances__muB0_0.5_keV}(d), the resonance condition can now be fulfilled for two different particle energies, for orbits belonging either in different families (upper branches) or in the same family (lower branch) where the latter presents a non-monotonic dependence on the energy. Moreover, the existence of a gap in the value range of $q_{kin}$ suggests that resonance conditions cannot be fulfilled for modes with $m/n$ ratio residing within this gap. For example, a perturbative mode with $(m, n)=(10,-3)$ that can resonate with the particle motion when $E_r=0$, cannot resonate when $E_r\neq 0$. Such features, evidently alter the resonant domain of the interaction upon the emergence of a radial electric field, cutting off or permitting specific modes to resonantly interact with the particles, indicating a major modification of stochastic particle transport in each case.

In Fig. \ref{fig:SpectrumResonances__muB0_2.0_keV}, orbital frequencies and kinetic-$q$ factor are depicted for low-energy particles corresponding to Figs. \ref{fig:phaseSpaceConfigurationSpace}(c), (d), with $\mu B_0=2.0\un{keV}$ and $P_\zeta=-0.025$. As already pointed out, the unconfined trapped orbits are transformed into confined trapped due to the presence of $E_r$, but more importantly, the kinetic-$q$ curve is significantly changed. 
The orbital spectrum is modified due to the presence of the radial electric field in such a way that resonances with modes having negative $m/n$ can take place, whereas these modes are non-resonant when $E_r=0$.

The case of mildly energetic counter-passing orbits corresponding to Figs. \ref{fig:phaseSpaceConfigurationSpace}(e)-(f) is shown in Fig. \ref{fig:SpectrumResonances__muB0_10.0_keV}. Although the modifications of the orbit shapes as well as of the orbital frequencies due to $E_r$ are not drastic, the respective effect on the kinetic-$q$ factor is quite important, as shown by comparing Fig. \ref{fig:SpectrumResonances__muB0_10.0_keV}(b) and (d). Besides the lack of resonances for $q_{kin}>2.62$ for $E_r \neq 0$, the non-monotonicity of the $q_{kin}(E)$ curve (also shown in the lower branch of Fig. \ref{fig:SpectrumResonances__muB0_0.5_keV}(d)) introduces two qualitatively different features in comparison to the $E_r=0$ case. 
First, the non-monotonicity indicates the likeliness of particles with different energies (located at different flux surfaces) to resonantly interact with the same perturbative mode. For instance, a mode with $(m,n)=(5,2)$ could resonantly affect orbits with energy $E=14.7\un{keV}$ when $E_r=0$, whereas all orbits with $E=\{14.7\un{keV}, 16.8\un{keV}, 18.4\un{keV}\}$ will be susceptible to resonantly interact with the specific mode, when a radial electric field is present. 
Second, the local extrema of the $q_{kin}(E)$ curve, where $q_{kin}'(E)=0$, indicate the existence of shearless points which have a great significance for stochastic transport, as they signify the onset of Stochastic Transport Barriers (STB), reducing the radial transport, even when chaotic orbits occur \citep{Morrison2000}, as will also be discussed in the following section.

\section{Resonant particle-mode interaction and stochastic transport in the presence of $\mathbf{E_r}$} \label{Resonant particle-mode interaction and stochastic transport in the presence of Er}

The calculations of the orbital spectrum and the kinetic-$q$ factor for the unperturbed GC motion in an axisymmetric LAR magnetic equilibrium in the presence of a radial electric field, allows for an a priori knowledge of important properties of stochastic transport under resonant particle-mode interaction with non-axisymmetric perturbative modes. More specifically, these calculations allow to pinpoint, with a remarkable accuracy, the regions of the six-dimensional phase space where significant particle and momentum transfer takes place, as well as the existence of dynamical barriers separating different regions of stochastic transport. This is of particular importance since the effects of resonant mode-particle interaction is strongly inhomogeneous in the phase space. It is worth emphasizing that this information does not require time-consuming numerical particle tracing and provides physical intuition on the characteristics of complex particle dynamics and stochastic transport. 

Time-independent perturbative magnetic modes, can be given in the form
\begin{equation}
    \delta \mathbf{B}=\nabla\times \alpha\mathbf{B}
\end{equation}
with $\alpha(\psi,\zeta,\theta)=\sum_{m,n}a_{mn}(\psi)\sin(n\zeta-m\theta)$, $a_{mn}(\psi)$ the amplitude profile of the perturbation,  $m,n$ integers that correspond to poloidal and toroidal harmonics of the mode, and $\psi=P_\theta$ for the case of a LAR equlibrium \citep{White2013b}. This type of magnetic field perturbation is particularly useful under the GC approximation as it sufficiently describes the $\nabla\psi$ component of any perturbation with practical interest, such as MHD activity of tearing or ideal modes, as well as Resonant Magnetic Perturbations (RMP) \citep{White2013a, White2013b}, whereas some possible limitations of that scheme are discussed in \citep{Ciaccio2013}. The perturbation term is directly introduced as a modification in normalized parallel velocity, yielding the expression of the perturbed GC Hamiltonian \citep{White1982, White1984}
\begin{equation}\label{eq:Perturbed Hamiltonian}
    \mathcal{H}=(\rho_\|-\alpha)^2 B^2/2 + \mu B + \Phi
\end{equation}

Without loss of generality, in terms of the resonant character of mode-particle interactions and their effects on stochastic transport, in the following analysis, we consider perturbations with a constant mode-amplitude, that is, $|a_{mn}|=\epsilon B_0$, with $\epsilon$ being designated as the ratio of the amplitude of the perturbative magnetic mode to the background magnetic field. It must be noted that, due to the nonlinear character of the canonical transformation to Action-Angle variables, a single mode in $(\zeta,\theta)$ results in a multi-mode Fourier expansion in terms of the $(\hat{\zeta},\hat{\theta})$ variables, according to Eq. \eqref{eq:Angles} \citep[p. 109]{White2014}.

The axisymmetry-breaking perturbative modes render the GC Hamiltonian non-integrable, since the $\zeta$-dependence of the modes results in a no longer invariant toroidal momentum $P_\zeta$, whereas the time-independent character of the perturbations ensures the invariance of the total energy $E$. These features suggest that GC dynamics under time-independent, non-axisymmetric perturbations take place on constant energy surfaces of the phase space and facilitate their study in terms of Poincar\'e surfaces of section in the  $(\zeta, P_\zeta)$ or $(\theta,P_\theta)$ planes, as well as the consideration of the kinetic-$q$ factor ($q_{kin}$) as a function of $P_\zeta$, for constant (invariant) values of the magnetic moment $\mu$ and the energy $E$.

\begin{figure}
     \centering
     \begin{subfigure}[t]{0.90\textwidth}
         \centering
         \includegraphics[width=\textwidth]{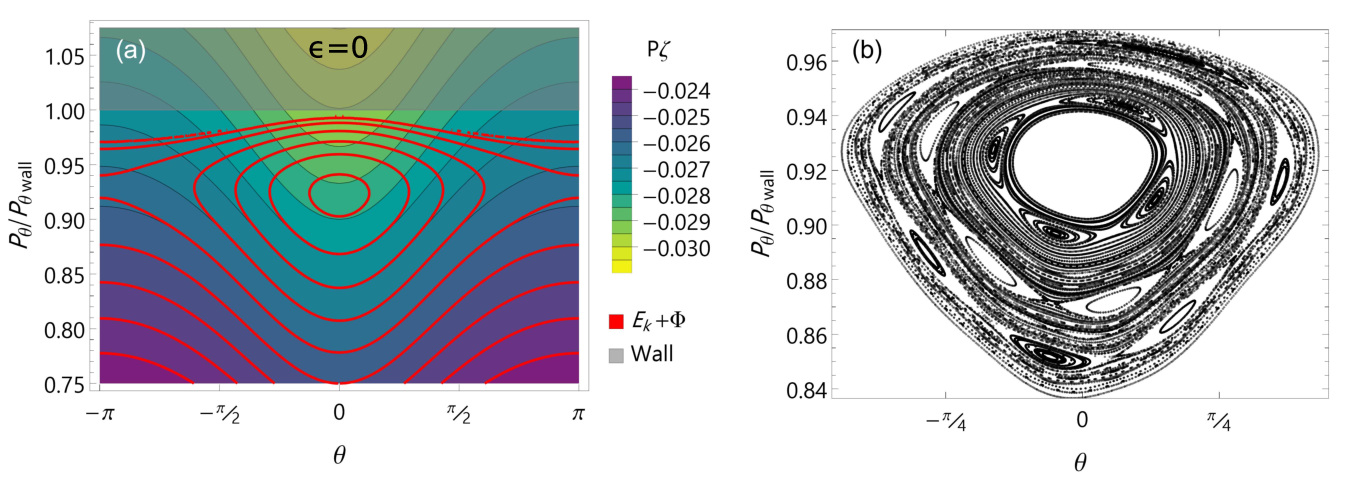}
         %\caption{}
         %\label{fig:}
     \end{subfigure}     
     \begin{subfigure}[t]{0.90\textwidth}
         \centering
         \includegraphics[width=\textwidth]{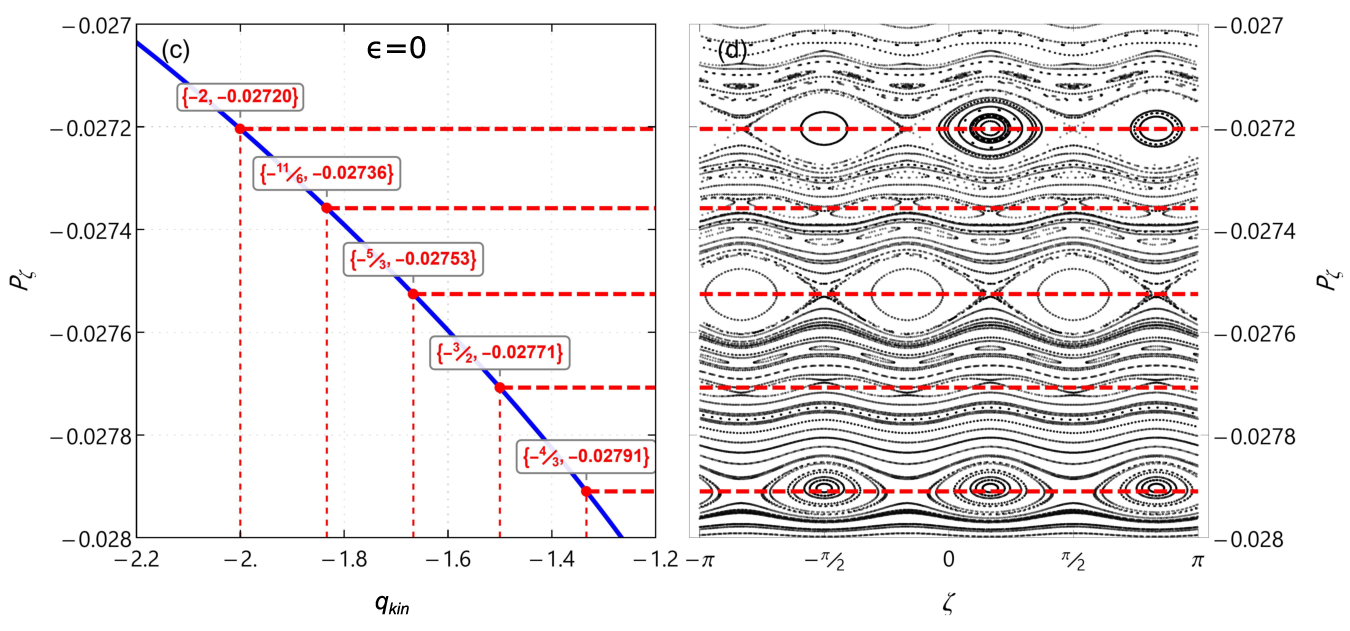}
         %\caption{$$}
         %\label{fig:}
     \end{subfigure}
        \caption{(a) Unperturbed phase space in $P_\zeta(\mu,E,P_\theta,\theta)$ representation for particles with $\mu B_0=0.5\un{keV}$, $E=1.5\un{keV}$. Coloured orbits are associated with the case $E_r=0$, while red orbits correspond to the $E_r\neq 0$ case. (b) Poincar\'e surface of sections at $\zeta=0$ for the perturbed Hamiltonian with $\epsilon=3\times10^{-6}B_0$ and $(m,n)=(5,-3)$, in $\theta-P_\theta$ cross-section. (c) Demonstration of the semi-analytical profile of kinetic-$q$ (drift-resonance ratio) characterizing constant energy orbits, for various values of $P_\zeta$, in the unperturbed system. (d) Poincar\'e surface of sections at $\theta=0$ when mode $(m,n)=(5,-3)$ is applied, in $\zeta-P_\zeta$ cross-section. Thick dashed red horizontal lines pinpoint the location of the excited resonant modes with respect to $P_\zeta$, exhibiting a perfect agreement between the numerically calculated values and the theoretically predicted values of the unperturbed system.}
        \label{fig:Poincare_muB0_0.5_keV}
\end{figure}

In Fig. \ref{fig:Poincare_muB0_0.5_keV} we consider the case of thermal particles with $\mu B_0=0.5\un{keV}$ and $E=1.5\un{keV}$. Panel (a) depicts the projection of the unperturbed orbits in the absence and the presence of a radial electric field, for different values of the canonical toroidal momentum $P_\zeta$. The nonzero $E_r$ introduces unperturbed trapped orbits (red closed orbits), not existing for $E_r=0$ (coloured orbits). The kinetic-$q$ factor under the presence of the radial electric field, determining the position of the resonances with the perturbative mode, is depicted in panel (c) as a function of $P_\zeta$. The labeled red points indicate rational values of $q_{kin}$ corresponding to fractions of small integers, and therefore primary resonances with different Fourier components (in terms of the Angle variables $\hat{\zeta}, \hat{\theta}$) of the perturbative mode with $(m,n)=(5,-3)$. Evidently, the exact locations of the resonant islands, as shown in a $(\zeta,P_\zeta)$ Poincar\'e surface of section (d), are obtained with a remarkable accuracy from the curve $q_{kin}(P_\zeta)$. The number of resonant islands in $(\zeta,P_\zeta)$ Poincar\'e surface of section coincides with the denominator of the resonant fractions, while the numerator corresponds to the number of islands in a $(\theta,P_\theta)$ Poincar\'e surface of section, as shown in panel (b). It should be mentioned that the island width corresponding to different resonances in both Poincar\'e surfaces of section depends on the amplitude of each Fourier component of the perturbative mode, as expressed in Angle variables, and determines the degree of the chaoticity of the phase space \citep{Chirikov1979, Lichtenberg1992}. In this specific case of thermal particles and perturbative mode, \sout{the part of} the phase space \sout{shown here} is mostly populated by regular orbits corresponding to well-defined KAM curves and island chains \citep{Lichtenberg1992}, and the perturbation does not lead to stochastic losses, since orbits are bounded by a surrounding KAM curve located inside the wall, as shown in panel (b). It is worth emphasizing that the application of this specific perturbative mode $(m, n)=(5,-3)$ is non-resonant in this domain of the phase space when $E_r=0$, as can be seen in the range of $q_{kin}$ values shown in Figure \ref{fig:SpectrumResonances__muB0_0.5_keV}(b). The corresponding Poincar\'e surface of section in this case simply consists of plain KAM lines.

\begin{figure}
     \centering
     \begin{subfigure}[t]{0.90\textwidth}
         \centering
         \includegraphics[width=\textwidth]{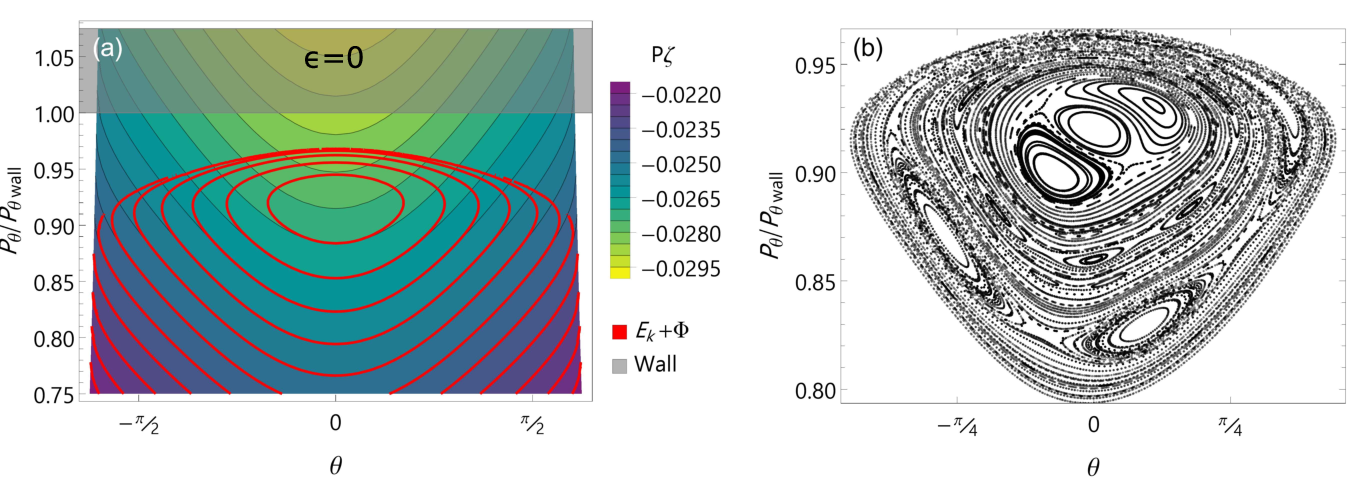}
         %\caption{}
         %\label{fig:}
     \end{subfigure}     
     \begin{subfigure}[t]{0.90\textwidth}
         \centering
         \includegraphics[width=\textwidth]{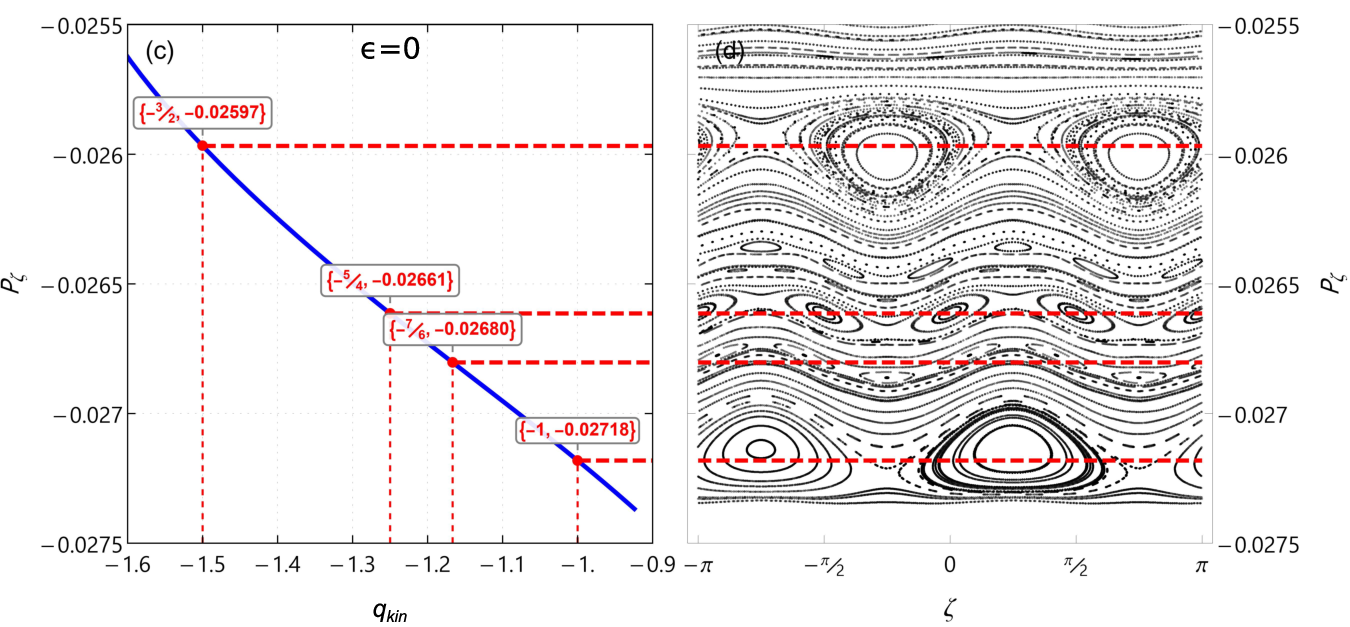}
         %\caption{$$}
         %\label{fig:}
     \end{subfigure}
        \caption{Same as figure \ref{fig:Poincare_muB0_0.5_keV}, for particles with $\mu B_0=2\un{keV}$, $E=2.2\un{keV}$. Applied perturbations correspond to the fluctuating magnetic mode $(m,n)=(3,-2)$ with $\epsilon=2\times 10^{-6}B_0$.}
        \label{fig:Poincare_muB0_2.0_keV}
\end{figure}

The case of low-energy particles with $\mu B_0=2\un{keV}$ and $E=2.2\un{keV}$ is shown in Figure \ref{fig:Poincare_muB0_2.0_keV}, where all unperturbed orbits are trapped for a nonzero $E_r$ (red orbits), as shown in panel (a). The location of resonance island chains, in the presence of a perturbative mode with $(m,n)=(3,-2)$, is accurately given by the rational values of $q_{kin}(P_\zeta)$ as shown in panels (c) and (d). Hence, the dominant mode $(m,n)=(3,-2)$ is identified by the two-island chain at $P_\zeta=-0.02597$ in panel (d) and by the three-island encircling chain in panel (c), having one of its elliptical points approximately at $\theta\simeq \upi/10$, $P_\theta\simeq 0.825 P_{\theta wall}$. Accordingly, the secondary resonance $(m,n)=(5,-4)$ is identified by the four-island chain at $P_\zeta=-0.0268$ and a corresponding five-island encircling chain with one of each elliptical points approximately at $\theta\simeq 0$, $P_\theta\simeq 0.86 P_{\theta wall}$. In both Poincar\'e surfaces of section, we can observe the onset of stochastization near the separatrix of the primary island with $(m,n)=(3,-2)$.

\begin{figure}
     \centering
     \begin{subfigure}[t]{0.90\textwidth}
         \centering
         \includegraphics[width=\textwidth]{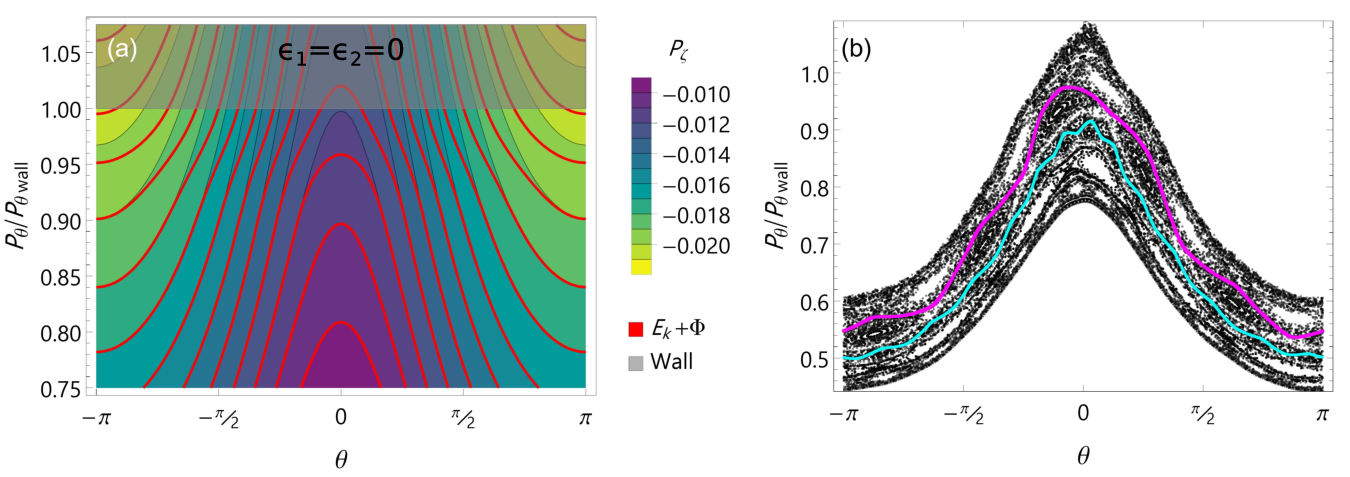}
         %\caption{}
         %\label{fig:}
     \end{subfigure}
     \begin{subfigure}[t]{0.90\textwidth}
         \centering
         \includegraphics[width=\textwidth]{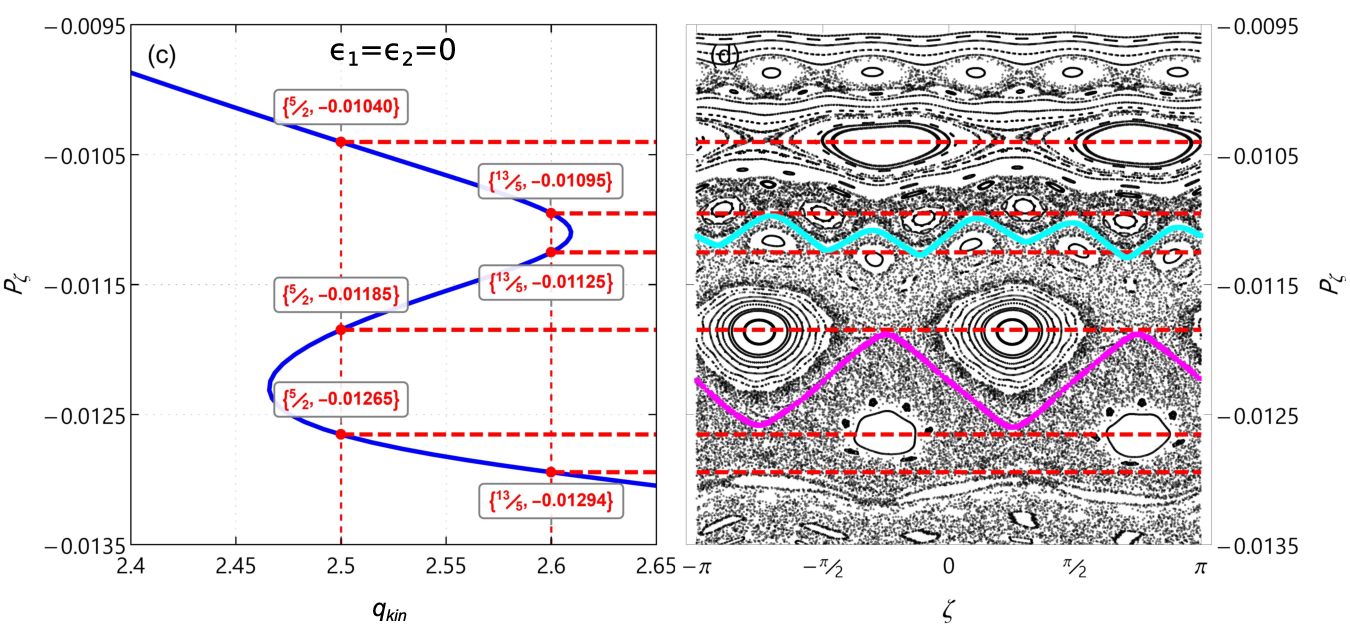}
         %\caption{$$}
         %\label{fig:}
     \end{subfigure}
        \caption{Same as figure \ref{fig:Poincare_muB0_0.5_keV}, for trajectories with $\mu B_0=10\un{keV}$, $E=18\un{keV}$. Applied perturbations correspond to the synergetic effect of two resonant modes $(m_1,n_1)=(5,2)$ with $\epsilon_1=5\times 10^{-6}B_0$, and $(m_2,n_2)=(13,5)$ with $\epsilon_2=0.9\times 10^{-6}B_0$. Magenta and cyan lines indicate the existence of two STBs, due to $E_r$ shear, effectively separating neighboring chaotic regions. The locations of the STBs are in perfect agreement with the analytically calculated locations of shearless points of $q_{kin}(P_\zeta)$ curve of (c).}
        \label{fig:Poincare_muB0_10.0_keV}
\end{figure}

For the case of higher-energy particles with $\mu B_0=10\un{keV}$ and $E=18\un{keV}$, depicted in Fig. \ref{fig:Poincare_muB0_10.0_keV}, it is evident that the presence of a radial electric field does not significantly modify the unperturbed counter-passing orbits, as shown in panel (a). However, in accordance to Fig. \ref{fig:SpectrumResonances__muB0_10.0_keV}, the kinetic-$q$ factor is drastically modified, as shown in panel (c). The non-monotonicity of the $q_{kin}(P_\zeta)$ curve results in the emergence of two local extrema, one local minimum at $(q_{kin},P_\zeta)_{min}=(2.47,-0.0123)$ and one local maximum at $(q_{kin},P_\zeta)_{max}=(2.61,-0.0111)$. The first consequence of this non-monotonic dependence concerns the multiplicity of specific resonances, in the sense that a particular resonant mode can interact with particles having different toroidal momentum $P_\zeta$, or equivalently, with particles that are located at different flux surfaces (different $P_\theta$). This feature is quite essential for particle and energy transport along the radial direction, as particles in different position can undergo a resonant interaction with the same perturbative mode. Provided that the perturbation strength is sufficiently large, in order for the resonances to become overlapping, this mechanism describes how a particle can start from the plasma core and drift all the way to the wall and become lost, entirely through stochastic transport. Second, the local extrema $q_{kin}'(P_\zeta)=0$ indicate regions of zero kinetic shear which are responsible for the generation of Stochastic Transport Barriers between two adjacent island-chains \citep{Horton1998, Morrison2000, Gobbin2011, Caldas2012, Marcus2019, Grime2023}. 

Figures \ref{fig:Poincare_muB0_10.0_keV}(b),(d) illustrate the Poincar\'e surfaces of section corresponding to the unperturbed cases of panels (a) and (c), at $\zeta=0$ and $\theta=0$, respectively, in the presence of two perturbing modes with $(m,n)=(5,2)$ and $(m,n)=(13,5)$ 
It is clearly shown that the locations of the resonant island chains are very accurately predicted by the values of the kinetic-$q$ factor $q_{kin}(P_\zeta)$. Resonances with the $(m,n)=(5,2)$ mode appear for three different values of $P_\zeta$, whereas resonances with the $(m,n)=(13,5)$ appear for two different values. In both cases, adjacent regions corresponding to resonances with the same mode, although highly chaotic, are well separated by persistent KAM curves forming Stochastic Transport Barriers and bounding the complex particle motion, at $P_\zeta$ values corresponding to local extrema of the kinetic-$q$ factor. It is worth emphasizing that non-axisymmetric perturbations induce magnetic field line chaoticity, with TBs bounding the chaotic regions of the magnetic field lines phase space, when non-monotonic $q$ profiles are considered. In such cases, STB appear for low-energy, field-line following particles at shearless points corresponding to local extrema of the $q$ factor, where $q'(\psi)=0$. However, for high-energy particle with large drifts across the magnetic field lines the effective kinetic shear is properly described by the kinetic-$q$ factor \citep{Gobbin2008, White2014, Antonenas2021}, with its extrema pinpointing the location of the STB. Both cases are introduced on equal footing in the third and fourth term, respectively, of the equation of motion for $\dot{\theta}$, shown in Eq. \eqref{eq:dtheta|dt}. Experimental findings indicate that there is a strong correlation between the appearance of local extrema in $q_{kin}$ and the elimination of transport for high-energy particles, when an edge-localized radial electric field is present \citep{Sanchis2019}.

\section{Summary and Conclusions} \label{sec:Conclusions}
The presence of a radial electric field in the pedestal area of a tokamak, imposes significant modification in the orbit topology. Rearranged or bifurcated critical points of the phase space may dramatically change the type of the orbits residing in that area and significantly modify particle prompt losses, especially for thermal or low-energy particles. The radial electric field significantly alters the orbital spectrum of the particles in such a way that it can modify the location of the resonances in the particle phase space, and prevent or allow resonances with specific modes. The resonance conditions are determined by the kinetic-$q$ factor, which contains all the essential information for the shear of the background magnetic field, the $\mathbf{E\times B}$ shear flow, as well as the neoclassical finite-orbit-width effects in a toroidal plasma configuration. Moreover, local extrema of the kinetic-$q$ factor correspond to locations where transport barriers are formed, preventing the extended stochastic particle transport. It is shown that the calculated kinetic-$q$ contains all the essential information of the macroscopic plasma configuration, allowing for the a priori knowledge of the exact locations of resonances and transport barriers, that determine particle, energy and momentum transport, as confirmed by numerical particle tracing simulations.

\section*{Acknowledgments}\label{sec:Acknowledgments}
Fruitful discussions with C. Karagianni are kindly acknowledged. The authors thank the anonymous referees for their valuable comments.

\section*{Funding}\label{sec:Funding}
This work has been carried out within the framework of the EUROfusion Consortium, funded by the European Union via the Euratom Research and Training Programme (Grant Agreement No 101052200 – EUROfusion). Views and opinions expressed are however those of the author(s) only and do not necessarily reflect those of the European Union or the European Commission. Neither the European Union nor the European Commission can be held responsible for them. The work has also been partially supported by the National Fusion Programme of the Hellenic Republic – General Secretariat for Research and Innovation. E.V. gratefully acknowledges funding from the European Research Council (ERC) under the European Union's Horizon 2020 research and innovation programme (grant agreement No. 805162).

\section*{Declaration of interest}\label{sec:Declaration_Interest}
The authors report no conflict of interest.

% ++++++++++++++++++++++++++++++++++++++++++++++++++++++++++++++++++++++++++++++++
\clearpage
\bibliographystyle{jpp}

\end{document}